# Saturation Magnetization of Inorganic/polymer Nanocomposites Higher than That of Their Inorganic Magnetic Component

**Yang Liu\*[1], Makoto Takafuji[2], Hirotaka Ihara[2],and Takeshi Wakiya[2]**
[1] *Beijing Key Lab of Special Elastomer Composite Materials, Department of Material Science and Engineering, Beijing Institute of Petrochemical Technology, 19 North Qingyuan Road, Beijing 102617, China. Tel: +86 10 81292129; \*Corresponding author: E-mail: blue_ocean3000@mail.dhu.edu.cn, yang.liu@bipt.edu.cn*
[2] *Department of Applied Chemistry and Biochemistry, Kumamoto University, Kumamoto 860-8555, Japan.*

**Abstract:** Herein, some magnetic nanoparticles (MNP)/clay/polymer nanocomposites have been prepared, whose saturation magnetization is higher than that of pure oleic acid coated MNP component. The existence of unique 'nano-network' structure and tight three-phase nano-interface in the nanocomposites contribute to the surprising saturation magnetization.

Inorganic/polymer composites are widely used in daily life because they exhibit both the functionality of inorganic component and the properties of polymer substrate. It is convenient to endow the composites functionalities by introducing different inorganic fillers, such as carbon black, carbon nanotube, graphene, clay, silica, quantum dot, polyhedral oligomeric silsesquioxane, magnetic particles, and so on[1-5]. As for the magnetic inorganic/polymer composites, the saturation magnetization of the composites is proportional to the amount of their magnetic components[6-8], lower than that of pure magnetic inorganic components, which is naturally believed to be true. However, the 'true' view is not right for all magnetic composites now: herein, some magnetic nanoparticles (MNP)/clay/polymer nanocomposites have been prepared, exhibiting ultrahigh saturation magnetization with low content of oleic acid coated MNP (OA-MNP) (i.e. 10 wt% OA-MNP with respect to the total weight), even stronger than that of pure OA-MNP component. The nanocomposite films were prepared by casting the mixture of clay (Laponite XLG), poly(butyl acrylate) (PBuA) emulsion, and OA-MNP. This is a breakthrough for the research of nanocomposites, which indicates that the functionality of the nanaocomposites can be better than that of their pure functional components by forming unique nanostructure in nanocomposites.

The preparation of the nanocomposites is shown in Scheme 1. First, the mixture of polymer emulsion and clay dispersion forms a system without aggregations. Then, the mixture turns into a gel during the drying process because of the formation of a "house of cards" of clay, as shown in Scheme 1b. This 'house of cards' results from the electrical attraction among the opposite charges of the surfaces and the edges of the clay platelets (i.e., negative charge and positive charge, respectively) (inset of Scheme 1b). At this stage, the mixture still contains a large amount of water; polymer particles and OA-MNP are separately fixed in the rooms of the house of cards. With further drying, when polymer particles are sufficiently close, coalescence among neighboring polymer particles takes place. Finally, all polymer particles coalesce into a macroscopic bulk film; clay platelets and OA-MNP are fixed in the polymer substrate. A unique 'nano-network' structure forms in the nanocomposites, shown in Scheme 1c. The obtained nanocomposite films are called B$x$G$y$M$z$, where B, G, and M stand for poly(butyl acrylate) (PBuA), Laponite XLG, and oleic acid coated magnetic nanoparticles (OA-MNP) respectively, and $x, y, z$ stands for PBuA, clay, and OA-MNP content (with respect to the total weight). Detailed composition for film casting is shown in Table 1. The casting process is similar to the casting process of clay/polymer composite latexes[9-11], except that OA-MNP is introduced as the third component. As shown in Scheme 1c, OA-MNP disperse in clay/polymer nanocomposites, simultaneously touching clay platelets and polymer.

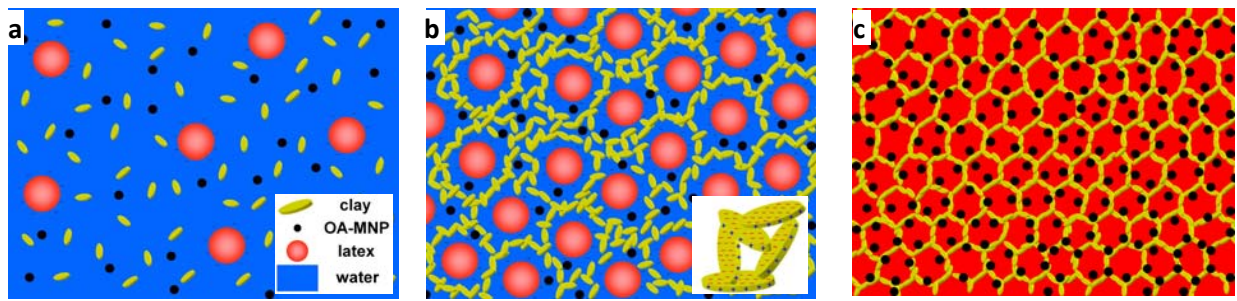

***Scheme 1.*** Preparation of magnetic nanoparticles (MNP)/clay/polymer nanocomposite films. a) Mixture of emulsion, clay aqueous dispersion, and oleic acid coated MNP (OA-MNP); b) Fixation of polymer nanospheres and MNP in the rooms of 'House of Cards'; c) Magnetic film containing 'nano-network' structure.

***Table 1.*** Composition for BxGyMz nanocomposite films

| Sample Name | Emulsion (PBuA, 21 wt%) | Clay Dispersion (2 wt%) | OA-MNP Slurry (4.2 wt%) | Water |
|---|---|---|---|---|
| | g | g | g | g |
| B80G10M10 | 1.1 | 1.5 | 0.714 | 4.686 |
| B60G30M10 | 0.829 | 4.5 | 0.714 | 1.957 |
| B40G50M10 | 0.553 | 7.5 | 0.714 | 0 |
| B80M10 | 3.3 | 0 | 2.142 | 2.6 |
| G50M10 | 0 | 3.75 | 0.357 | 4 |

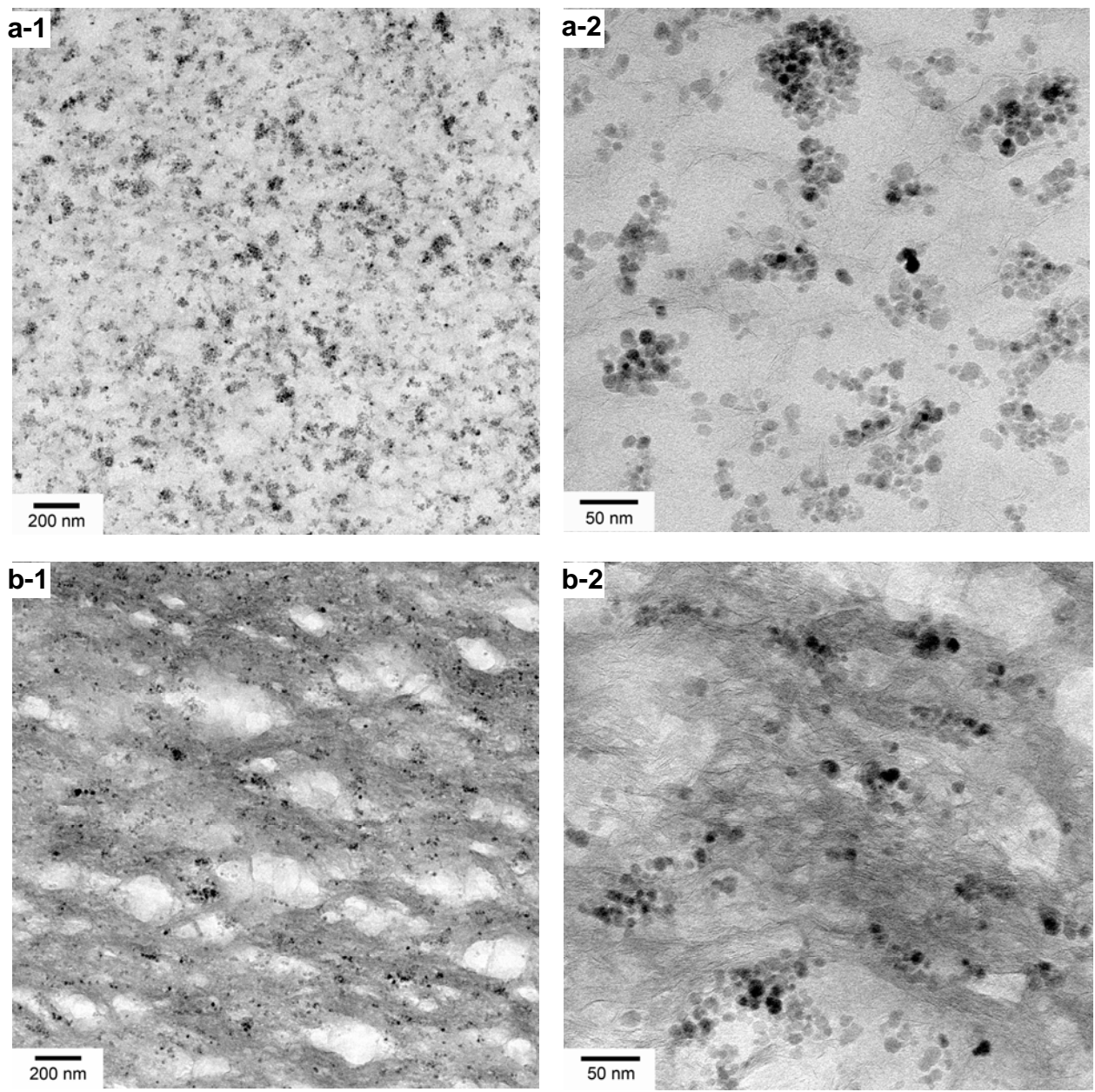

**Figure 1.** TEM photos of BxGyMz. a) B80G10M10; b) B40G50M10.

TEM photos confirm the existence of the nano-network structure, shown in Figure 1 (or Figure S2: high resolution photos of TEM), where intercalated clay platelets play frames of the network and divide the nanocomposites into many nano-polyhedrons filled with polymers (i.e. PBuA). With increasing clay content, the size ($\zeta$) of the nano-polyhedrons decreases and the frame thickness (D) increases: $\zeta$ = 80–200 nm and D = 5–30 nm for B80G10M10, and $\zeta$ = 60–100 nm and D = 10–400 nm for B40G50M10. There are OA-MNP aggregates in both B80G10M10 and B40G50M10. Nano-interfaces of three phases around the OA-MNP aggregates are found in the nanocomposites. The three phases are clay platelets, MNP particles and polymers. The nano-interfaces are believed to be tight because of the merging force of PBuA latexes during the casting process.

B$x$G$y$M$z$ nanocomposite films with such 'nano-network' structure and tight three-phase nano-interface exhibit surprising ultrahigh saturation magnetization, even higher than that of pure OA-MNP (i.e. 100 emu/g for B80G10M10, and 30 emu/g for pure OA-MNP), while the saturation magnetization of other two nanocomposites - B80M10 (the mixture of PBuA and OA-MNP) and G50M10 (the mixture of Laponite XLG and OA-MNP) is much lower than that of pure OA-MNP, shown in Figure 2. There is no chemical reaction and no new component during the casting process, which is confirmed by the FTIR (Figure 3), so the surprising saturation magnetization results from the physical structure in B$x$G$y$M$z$ films. Since the unique 'nano-network' structure and the tight three-phase nano-interface, confirmed by TEM photos, can exist only in the B$x$G$y$M$z$ nanocomposites instead of the two-component nanocomposites (i.e. B80M10 and G50M10). Therefore, the 'nano-network' structure and the tight three-phase nano-interface is the reason for the surprising saturation magnetization. In addition, the saturation magnetization decreases with increasing clay content: the saturation magnetization of B40G50M10 is lower than that of B80G10M10, shown in Figure 2. The OA-MNP aggregates in B80G10M10 disperse uniformly in the polymer substrate, generally near the plates of the nano-polyhydrons and surrounded by polymer and clay platelets; while most OA-MNP aggregates in B40G50M10 disperse in the frame of the 'nano-network', surrounded by more clay platelets and less polymer, shown in Figure 1b, which may make the nano-interface looser because of the decreasing number of polymer chains. The looser three-phase nano-interfaces probably give rise to the decrease in saturation magnetization.

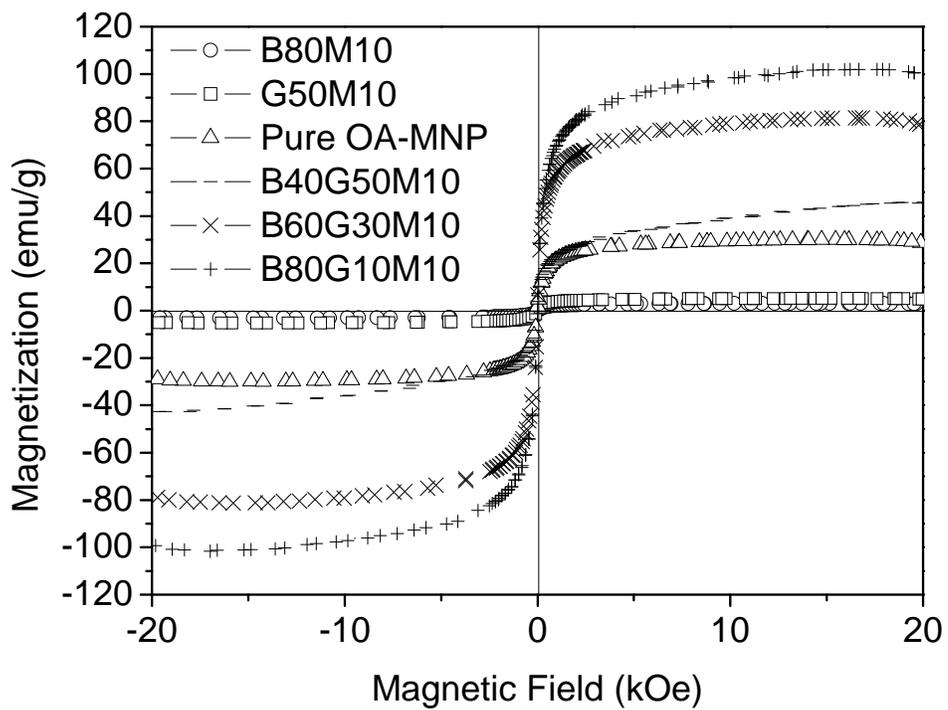

***Figure 2***. Magnetic hysteresis loop of BxGyMz, B80M10, G50M10, and pure OA-MNP.

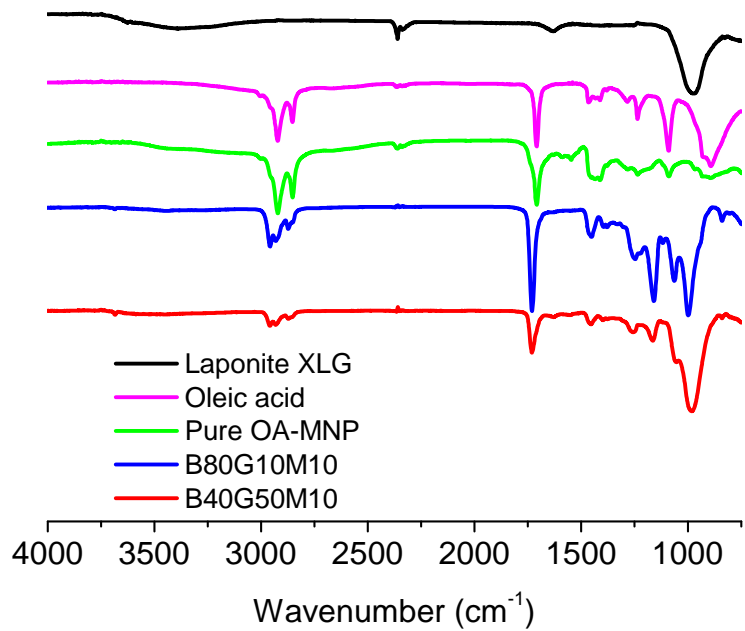

***Figure 3***. FTIR of BxGyMz, oleic acid, OA-MNP and laponite XLG.

In conclusion, some magnetic nanoparticles (MNP) /clay/polymer nanocomposites have been prepared, whose saturation magnetization is higher than that of pure oleic acid coated MNP component. The existence of unique 'nano-network' structure and tight three-phase nano-interface in the nanocomposites contribute to the surprising saturation magnetization. Although the detailed theoretical explanation for the surprising saturation magnetization in the nanocomposites needs to be further investigated in future, the experimental discovery is believed to greatly promote both experimental and theoretical research in magnetism of nanocomposites and to widen the applications of magnetic nanocomposites. And perhaps the unique nanostructure will also improve the functionality of other functional clay/polymer nanocomposites, like quantum dot/clay/polymer nanocomposites etc.

**Experimental**

**Materials:** Butyl acrylate (BuA) (98%, Wako Co., Japan), Laponite XLG (Rockwood Co., U.S., $Mg_{5.34}Li_{0.66}Si_8O_{20}(OH)_4Na_{0.66}$), sodium dodecyl sulfate (SDS) (99%, Nacalai Tesque Inc., Japan), ammonium persulfate (APS) (98%, Kanto Chemical Co., Inc., Japan). BuA were used after removing inhibitors. Other reagents were used as received.

**Preparation of Polymer Emulsion:** The emulsion was prepared by conventional emulsion polymerization using monomer (15.6 g; BuA), SDS (0.15 g) as the surfactant, APS (0.06 g) as the initiator and water (57 g). The reaction was run at 80 $^o$C for 8 h under nitrogen atmosphere. The average diameter of PBuA latex is 67.2 nm.

**Preparation of oleic acid coated magnetic nanoparticles (OA-MNP):** Preparation of OA-MNP nanoparticles was carried out according to the well-established co-precipitative reaction protocol. Typically, $FeCl_2\cdot4H_2O$ (2.35 g) and $FeCl_3\cdot6H_2O$ (0.86 g) were dissolved in water (40 mL), and co-precipitated by adding concentrated ammonia (28%, 5 mL) under $N_2$ atmosphere at 60 $^o$C. Oleic acid (1g) was slowly dropped into the reactor under vigorous stirring. The dispersion was heated to 90 $^o$C, and was kept at 90 $^o$C for 30 min. Then, it was cooled to room temperature. The OA-MNP was poured into a dialysized bag and was dialyzed with excessive distilled water until the pH of the OA-MNP dispersion dropped to 7. The obtained OA-MNP slurry was stored for use. The content of OA-MNP and MNP are 4.2 wt% and 2.1 wt%, respectively, shown in Figure S1.

**Preparation of BxGyMz Nanocomposite Films:** First, clay (0.6 g) was added into water (30 g) under stirring for 2 h to obtain a transparent aqueous dispersion. Then, an amount of the dispersion (1.5–7.5 g) was diluted by water (0–4.686 g). The polymer emulsion (0.553–1.1 g) and the OA-MNP slurry (0.714 g) were added into the diluted clay dispersion, respectively. Finally, the mixture was poured into a polyethylene container and dried in an oven at 50 $^{o}$C for 30 h. For all samples, the total solid content was fixated at 0.3 g. Here, the samples are expressed as B$x$G$y$M$z$, where B, G, and M stand for poly(butyl acrylate) (PBuA), Laponite XLG, and modified magnetic nanoparticles (OA-MNP) respectively, and $x, y, z$ stands for PBuA, clay, and OA-MNP content (with respect to the total weight).

**Measurement methods:** Transmission electron microscopy (TEM) (JEOL JEM2100): 50 nm thin film, 100 kV. Thermogravimetric analysis was performed on a Seiko Exstar 6000 TG/DTA 6200 thermal analyzer (Seiko Instruments, Chiba, Japan) in static air from 30 to 600 $^{o}$C with a heating rate of 10$^{o}$C min$^{-1}$. Vibrating sample magnetometer (VSM, Tamakawa Co., Japan) was calibrated by nickel sheet. Nanocomposite films were cut as 5mm×10mm rectangles (thickness: 40-100 μm; weight: 2.8-7 mg) for measurement. Fourier transform IR (FTIR): 40-100 μm film, 400-4000 cm$^{-1}$.

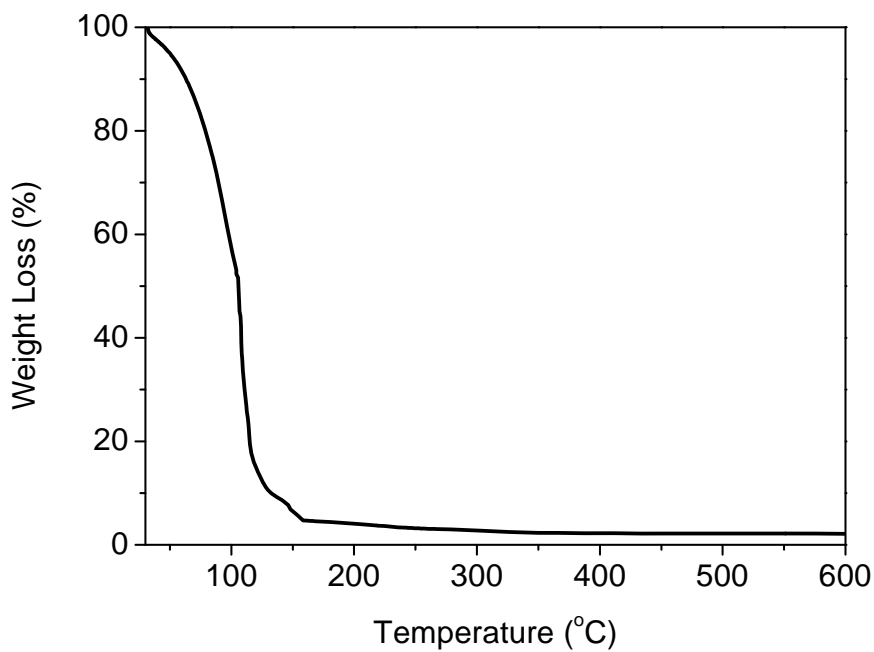

Figure S1. Thermogravity graph of OA-MNP slurry.

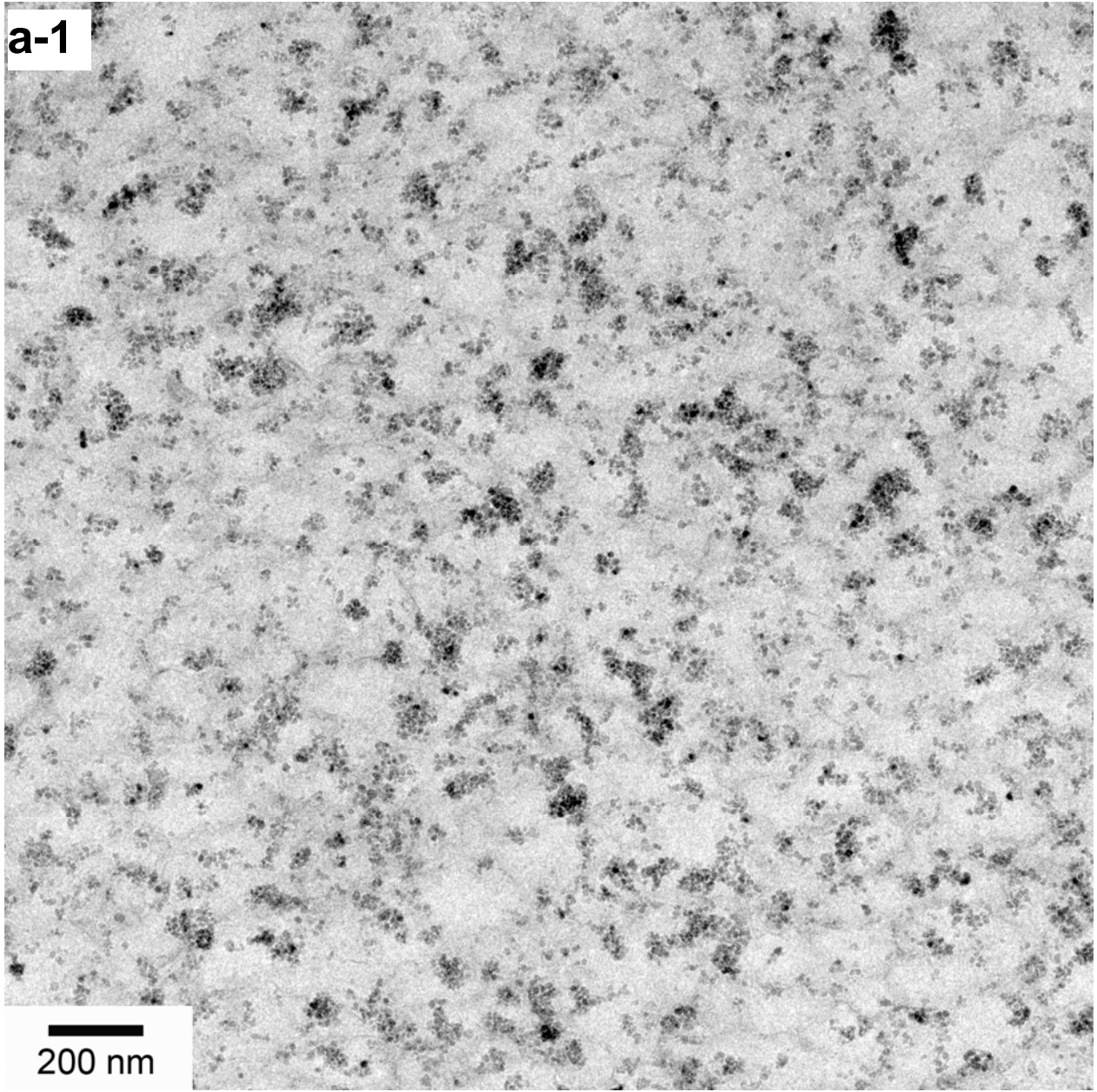

a-1

200 nm

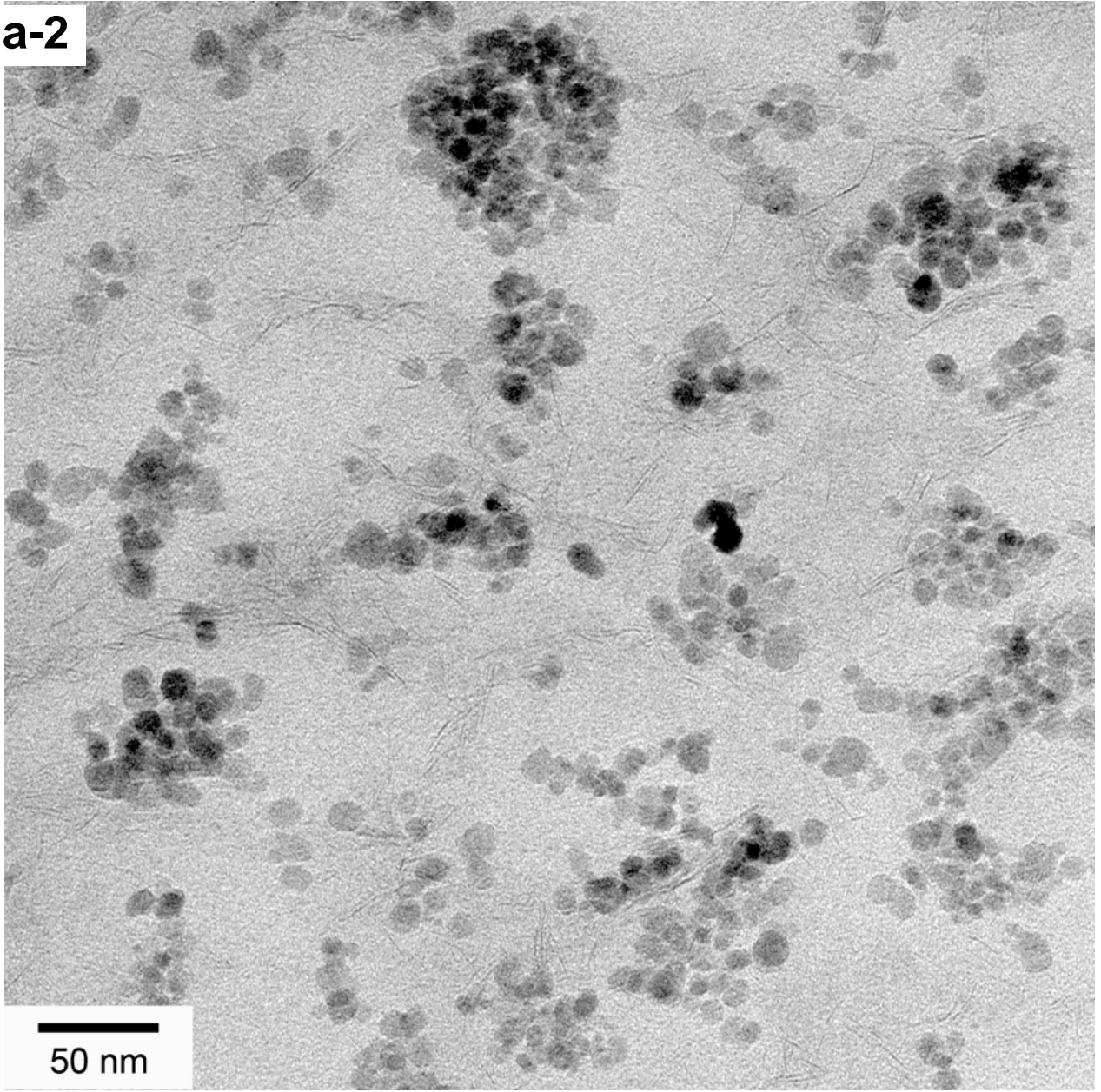

a-2

50 nm

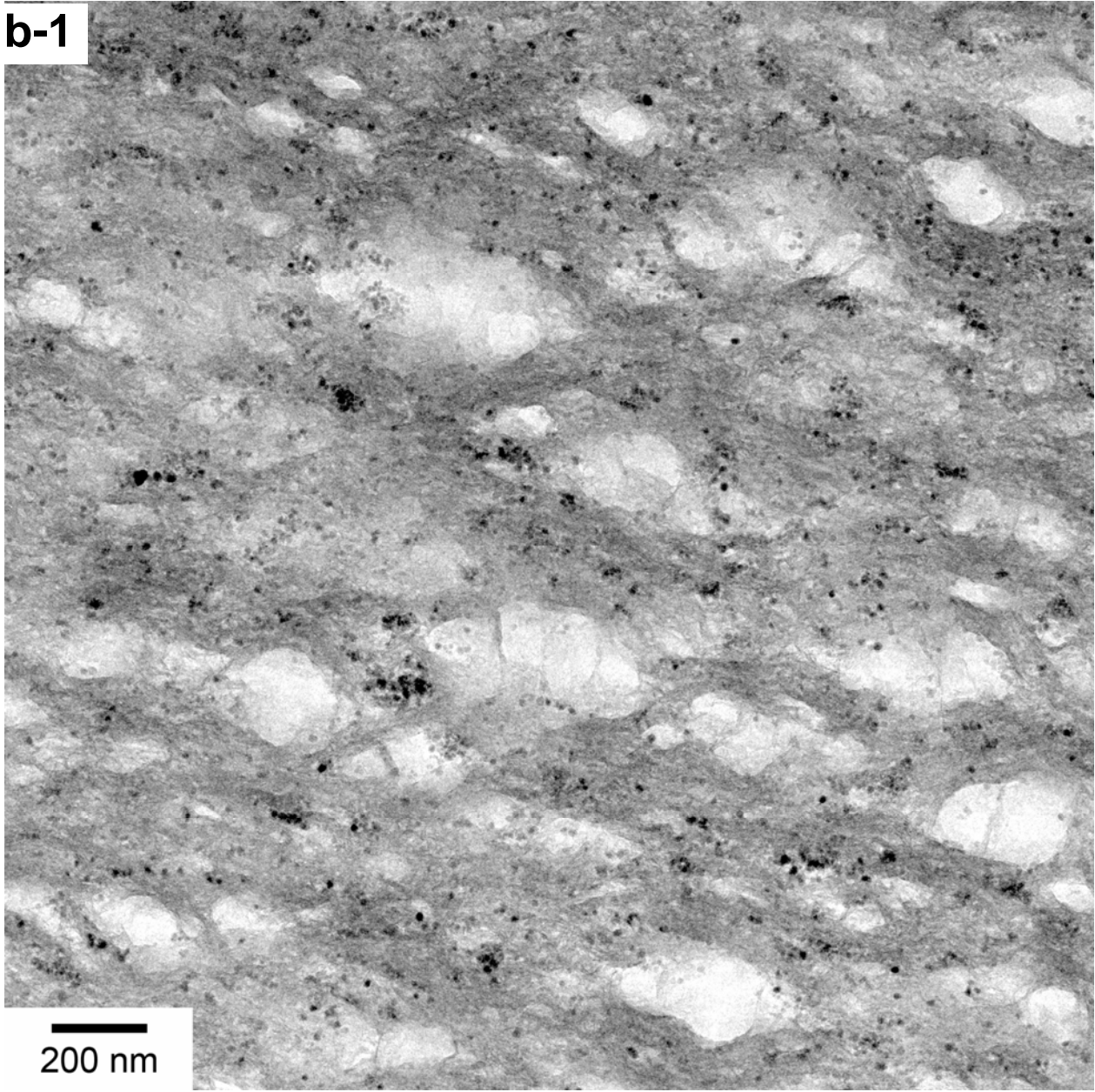

**b-1**

200 nm

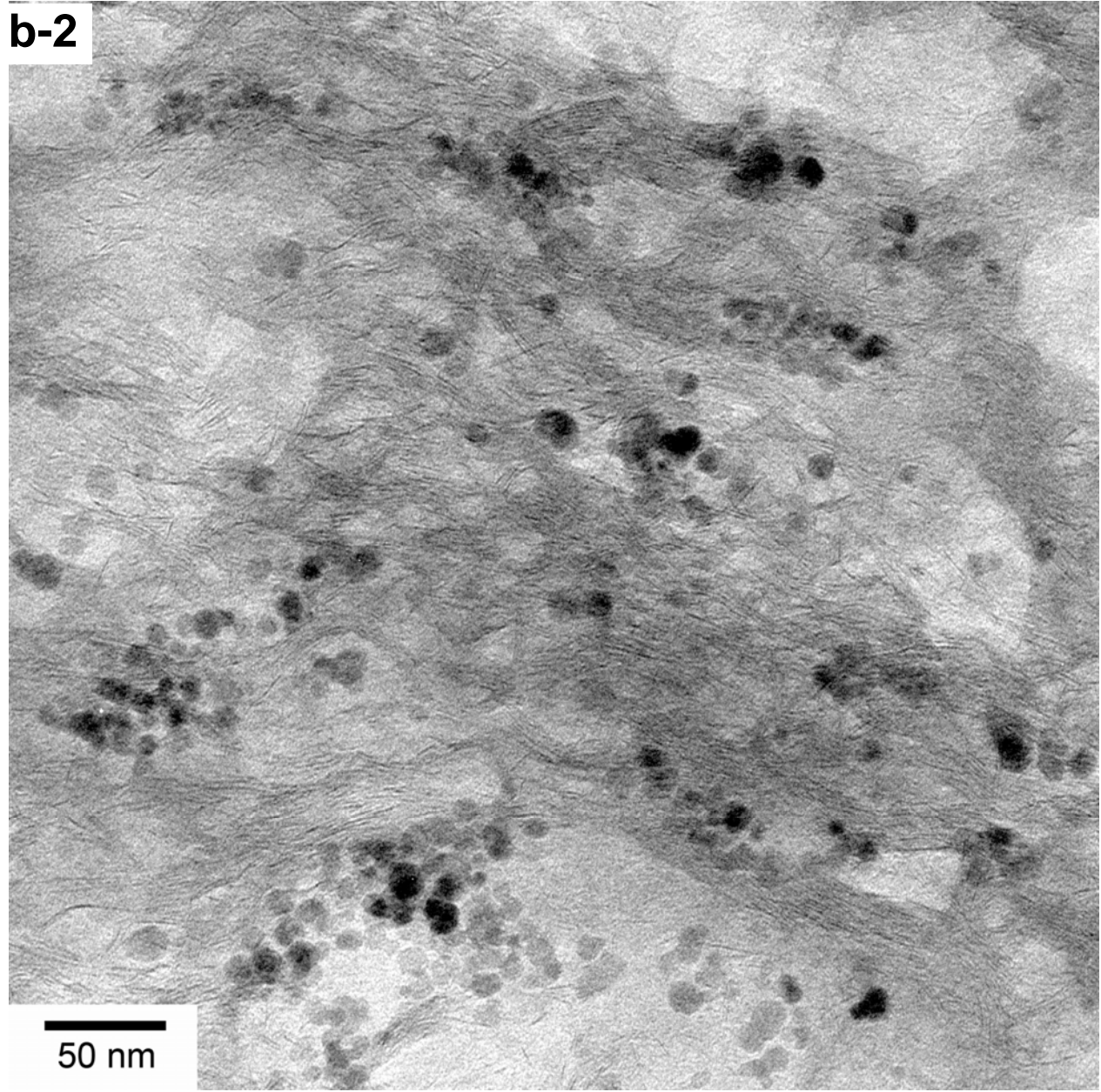

Figure S2. High resolution photos of TEM of B$x$G$y$M$z$

**Acknowledgements**

This research is financially supported by Beijing Natural Science Foundation(No.2122015), and the Japan Society for the Promotion of Science for Foreign Researchers (P08043).

**Notes and references**